\documentclass[jcappub,11pt,a4paper]{article}

\usepackage{jcappub}

\usepackage{graphicx}
\usepackage[export]{adjustbox}

\usepackage{url}
\usepackage{hyperref}
\usepackage{hypernat}

\usepackage{slashed}

\usepackage{layouts}

\usepackage[caption=false,labelfont=normalsize,labelformat=empty,textfont=normalsize,justification=centering]{subfig}

\renewcommand{\Re}{\mathrm{Re}}
\renewcommand{\Im}{\mathrm{Im}}

\newcommand{\iu}{\mathrm{i}}

\newcommand{\Tr}{\mathrm{Tr}}

\begin{document}

\title{Mass-derivative relations and unitarity constraints for CP asymmetries at finite temperature}

\author[a]{Tom\'a\v s Bla\v zek,}
\author[a,1]{Peter Mat\'ak,\note{Corresponding author.}}
\author[a]{Viktor Zaujec}
\emailAdd{peter.matak@fmph.uniba.sk}
\affiliation[a]{Department of Theoretical Physics, Comenius University in Bratislava,\\ Mlynsk\'a dolina, 84248 Bratislava, Slovak Republic}

\abstract{
Within the seesaw type-I leptogenesis, we formulate $CPT$ and unitarity constraints for the equilibrium reaction rate $CP$ asymmetries and consider thermal mass and quantum statistics. We demonstrate that including higher-order perturbative corrections in the classical Boltzmann equation remarkably induces quantum effects into the kinetic theory.
}

\keywords{leptogenesis, baryon asymmetry, physics of the early universe}

\maketitle

\section{Introduction}\label{sec1}

The conditions for matter asymmetry generation, formulated by Sakharov, require the mat\-ter in the early universe to be out of thermal equilibrium \cite{Sakharov:1967dj}.
For a specific field-theoretical model, the $CPT$ and unitarity constraints provide a perturbative mechanism to check that no asymmetry is generated in equilibrium \cite{Dolgov:1979mz,Kolb:1979qa}. Theoretical development and related diagrammatic formalism were focused mainly on using the classical kinetic theory and Boltzmann equation \cite{Roulet:1997xa, Bhattacharya:2011sy, Baldes:2014gca, Baldes:2014rda, Baldes:2015lka}. Quantum statistics has been considered for the leading-order asymmetries only \cite{Kolb:1979qa, Nanopoulos:1979gx, Hook:2011tk, Blazek:2021zoj}. Using an unconventional diagrammatic approach introduced in Ref. \cite{Blazek:2021zoj, Blazek:2021gmw}, we formulate $CPT$ and unitarity constraints for higher-order asymmetries, including thermal-mass effects in the leading-order asymmetry kinematics.

In the rest of this section, we introduce two ingredients that we find particularly useful for early universe calculations. First, we explain the role of perturbative unitarity in reaching completeness in the quantum description of particles' interactions. Next, the use of cylindrical diagrams introduced in our previous work in Ref. \cite{Blazek:2021zoj}, to account for quantumness in the kinetic theory of a many-body system, is briefly reviewed. In Sect. \ref{sec2}, the seesaw type-I leptogenesis model and top-Yukawa corrections to the leading-order asymmetries are considered. This study is not primarily focused on lepton number asymmetry generation and the model is only a working example. The anomalous thresholds in the relevant forward-scattering diagrams are shown to be related to the Higgs thermal mass. Finally, in Sect. \ref{sec3}, we demonstrate that taking all possible holomorphic cuts into account, including windings of propagators, while using zero-temperature rates, results in quantum thermal effects showing up in the unitarity relations.

\subsection{Unitarity and holomorphic cuts}

Before we dive into the kinetic description of the matter in the early universe, let us comment on the perturbative treatment of interactions by means of zero-temperature quantum field theory. Scattering matrix elements for any $i\rightarrow f$ reaction can be split as $S_{fi}=\delta_{fi}+\iu T_{fi}$. The transition probability per unit volume per unit time then equals
\begin{align}\label{eq1.1}
\frac{1}{V_4}\vert T_{fi}\vert^2=(2\pi)^4\delta^{(4)}(p_f-p_i)\vert M_{fi}\vert^2.
\end{align}
The amplitude squared can also be expressed as a product of two diagrams
\begin{align}\label{eq1.2}
-\iu M^\dagger_{if} \iu M^{\vphantom{\dagger}}_{fi}
\end{align}
that, in turn, can be interpreted as a Cutkosky cut of a forward-scattering diagram. It is important to note that to achieve the asymmetry cancelation when summing over the final states, and also to obtain infrared-finite results through the Kinoshita-Lee-Nauenberg theorem \cite{Kinoshita:1962ur, Lee:1964is, Frye:2018xjj}, for each initial state, one needs to consider a complete set of the final states at a given perturbative order. These final states follow from the diagrammatic structure of the theory. Starting with forward-scattering diagrams for an initial state $i$, one performs all their kinematically-allowed cuts. However, the standard Cutkosky cuts include complex conjugated diagrams, as seen in Eq. \eqref{eq1.2}. It turns out to be very convenient to employ the so-called \emph{holomorphic cutting rules} instead \cite{Coster:1970jy, Bourjaily:2020wvq, Hannesdottir:2022bmo, Blazek:2021olf}. Using the $S$-matrix unitarity while expanding $S^\dagger = S^{-1}$ into a geometric series, we obtain
\begin{align}\label{eq1.3}
\vert T^{\vphantom{\dagger}}_{fi}\vert^2 = 
-\iu T^{\vphantom{\dagger}}_{if} \iu T^{\vphantom{\dagger}}_{fi} + \sum_n \iu T^{\vphantom{\dagger}}_{in\vphantom{f}}\iu T^{\vphantom{\dagger}}_{nf}\iu T^{\vphantom{\dagger}}_{fi} - \ldots
\end{align}
Therefore, instead of the single Cutkosky cut with complex conjugation, we may cut the diagrams multiple times with appropriate signs according to Eq. \eqref{eq1.3}, simply putting the intermediate states on their mass shell. We emphasize that considering all kinematically-allowed cuts, one often has to deal with \emph{anomalous thresholds} \cite{Mandelstam:1960zz, Cutkosky:1961, Goddard:1969ci, Hannesdottir:2022bmo}. At low perturbative orders, these typically appear in the presence of unstable particles, such as right-handed neutrinos of the seesaw type-I leptogenesis. In the square of the amplitude, they emerge as interference between connected and disconnected diagrams.

Theories with irreducible complex phases in couplings allow for $CP$ violating in\-te\-ra\-ctions. Employing the unitarity expansion of Eq. \eqref{eq1.3}, we obtain for the $i\rightarrow f$ reaction asymmetry \cite{Blazek:2021olf}
\begin{align}\label{eq1.4}
\Delta \vert T_{fi}\vert^2 =\vert T_{fi}\vert^2-\vert T_{if}\vert^2= 
&\sum_{n}(\iu T_{in} \iu T_{nf} \iu T_{fi} - \iu T_{if} \iu T_{fn} \iu T_{ni})\\
&-\sum_{n,m}(\iu T_{in} \iu T_{nm} \iu T_{mf} \iu T_{fi} - \iu T_{if} \iu T_{fm} \iu T_{mn} \iu T_{ni})\nonumber\\
&+\vphantom{\sum_{n}}\ldots\nonumber
\end{align}
where $CPT$ symmetry has been assumed. Note that a non-vanishing asymmetry is obtained for forward-scattering diagrams with at least two simultaneous kinematically-allowed cuts. Usually, this condition is phrased in terms of the imaginary part of loop integrals. Furthermore, the two terms in each bracket only differ when complex couplings are present. Even at this level, we may appreciate the immediate insight from Eq. \eqref{eq1.4} generalizing the approach of Ref. \cite{Roulet:1997xa} to higher-order asymmetries. When summing over the final states $f$, the terms in the brackets cancel pairwise and the sum of the asymmetries vanishes explicitly. Therefore, the $CPT$ and unitarity constraints -- the main objective of this work -- are manifested in a simple diagrammatic way.

\subsection{Kinetic theory and cylindrical diagrammatic representation}

To study particle densities and their asymmetries in the early universe, the Boltzmann equation is often used as an approximation. Counting the number of the $i\rightarrow f$ reactions in a unit volume and unit time enters the computation through the circled rate
\begin{align}\label{eq1.5}
\mathring{\gamma}_{fi}=\frac{1}{V_4}
\int\prod_{\forall i} [d\mathbf{p}_i] \mathring{f}_i(p_i) 
\int\prod_{\forall f}[d\mathbf{p}_f] 
\bigg(-\iu T_{if}\iu T_{fi} + \sum_n \iu T_{in\vphantom{f}}\iu T_{nf} \iu T_{fi}+\ldots\bigg)
\end{align}
with
\begin{align}\label{eq1.6}
[d\mathbf{p}_i]=\frac{d^3\mathbf{p}_i}{(2\pi)^3 2E_i}
\end{align}
denoting the Lorentz-invariant measure in the momentum space. The circle refers to the use of zero-temperature quantum field theory, describing the interactions as if appearing in vacuum. In thermal equilibrium, the circled particle phase-space density $\mathring{f}_i$ corresponds to the Maxwell-Boltzmann exponential. 

How can we properly account for quantum effects in the above rates, going beyond the classical description represented by the Boltzmann equation? Naively, one may put the quantum statistical factors for the final states by hand. However, in higher-order calculations, a modification of the on-shell part of the propagators occurs. For a consistent treatment of such effects, and also to avoid double-counting of the on-shell intermediate states, we have to consider the evolution of the density matrix. This is usually achieved in a top-down approach \cite{Prokopec:2003pj, Prokopec:2004ic, DeSimone:2007gkc, Biondini:2017rpb}, starting with the Keldysh-Schwinger formalism \cite{Schwinger:1960qe, Keldysh:1964ud} reduced to the Kadanoff-Baym kinetic equations \cite{Baym:1961zz}. Our approach is different. In generality we may write the density matrix as \cite{Wagner:1991}
\begin{align}\label{eq1.7}
\hat{\rho}=\frac{1}{Z}\exp{\big\{-\hat{F}\big\}},\quad\text{where}\quad
Z=\Tr\exp{\big\{-\hat{F}\big\}}
\end{align}
for a Hermitian operator $\hat{F}$. Neglecting multiparticle correlations, we may also write
\begin{align}\label{eq1.8}
\hat{F} = \sum_p F^{\vphantom{\dagger}}_p a^\dagger_p a^{\vphantom{\dagger}}_p  
\end{align}
where $p$ labels a single-particle state, while $a^\dagger_p$ and $ a^{\vphantom{\dagger}}_p$ denote creation and annihilation operators, respectively. Then for the mean occupation number, we obtain
\begin{align}\label{eq1.9}
f(p) = \Tr\big\{\hat{\rho}a^\dagger_p a^{\vphantom{\dagger}}_p\big\}
=\frac{1}{\exp{\big\{F_p\big\}}\mp 1}
\end{align}
where the upper (lower) sign corresponds to the bosonic (fermionic) particle species. In analogy to the equilibrium case, we define a circled density \cite{Blazek:2021zoj}
\begin{align}\label{eq1.10}
\mathring{f}(p)=\exp{\big\{-F_p\big\}} = \frac{f(p)}{1\pm f(p)}.
\end{align}
We further assume the mean free path of particles is much bigger than the range of their interactions. Then we may use the asymptotic states necessary to describe the system's temporal evolution in terms of the $S$-matrix elements \cite{McKellar:1992ja}
\begin{align}\label{eq1.11}
\hat{\rho}\quad\rightarrow\quad \hat{S}\hat{\rho}\hat{S}^\dagger.
\end{align}
This includes changes in higher occupation numbers represented by quantum or uncircled rates. While using the same $S$-matrix elements as in Eq. \eqref{eq1.5}, the uncircled $\gamma_{fi}$ is obtained from cuttings of diagrams with propagators wound up on a cylindrical surface (see Ref. \cite{Blazek:2021zoj} for more details). For each line entering the initial state, one power of the respective circled density is added. Summing over all possible winding numbers, quantum densities are obtained from the circled ones in the geometric series
\begin{align}\label{eq1.12}
f(p) = \sum^\infty_{w=1}(\pm 1)^{w-1}\mathring{f}(p)^w.
\end{align}
Similarly, for each intermediate-state cut in Eq. \eqref{eq1.5}, a quantum statistical factor $1\pm f(p)$ is obtained from the summation over the winding numbers.

In the rest of this work, we stick to the equilibrium phase-space densities, as they are relevant to the asymmetry cancelations. The detailed balance principle then implies
\begin{align}\label{eq1.13}
\prod_{\forall i}\mathring{f}_i(p_i) = \prod_{\forall f}\mathring{f}_f(p_f)
\end{align}
and, using Eqs. \eqref{eq1.4} and \eqref{eq1.5}
\begin{align}\label{eq1.14}
\sum_f\Delta\mathring{\gamma}_{fi}=0.
\end{align}
Our main goal is to generalize this relation, following from unitarity and $CPT$ symmetry, for the uncircled or quantum-corrected rates at higher perturbative orders.

\section{Anomalous thresholds and thermal mass}\label{sec2}

In this section, we consider the seesaw type-I leptogenesis to study the connection between higher-order asymmetries and Higgs thermal-mass effects in the leading-order asymmetry ki\-ne\-ma\-tics. From the standard model part, we include the right- and left-handed top quark fields, $t$ and $Q$, respectively, and their Yukawa interactions. In addition to that, the Lagrangian density
\begin{align}\label{eq2.1}
\mathcal{L} \supset -\frac{1}{2} M_i \bar{N}_i N_i - (Y_{\alpha i} \bar{N}_i P_L l_\alpha H +
Y_t \bar{t} P_L Q H + \mathrm{H.c.}).
\end{align}
contains the interactions of right-handed neutrinos $N_i$ and their Majorana masses. The couplings $Y_{\alpha i}$ may come with irreducible complex phases responsible for the lepton number asymmetry generation \cite{Fukugita:1986hr,Covi:1996wh,Buchmuller:1997yu}. The role of the top-Yukawa interactions in leptogenesis has been studied in the literature within the classical kinetic theory \cite{Luty:1992un, Pilaftsis:2003gt, Pilaftsis:2005rv,Buchmuller:2004nz, Abada:2006ea, Nardi:2007jp, Davidson:2008bu, Racker:2018tzw}, thermal field theory \cite{Giudice:2003jh, Salvio:2011sf, Bodeker:2017deo}, and non-relativistic effective theory \cite{Biondini:2013xua, Biondini:2015gyw, Biondini:2016arl}. To our best knowledge, $CPT$ and unitarity constraints for higher-order asymmetries have only been formulated for the classical case\footnote{Within the closed-time-path formalism the leading-order asymmetry cancelations have been considered in Ref. \cite{Garbrecht:2013iga}.}.

\begin{figure}
\subfloat{\label{fig1a}}
\subfloat{\label{fig1b}}
\subfloat{\label{fig1c}}
\subfloat{\label{fig1d}}
\centering
\includegraphics[scale=1]{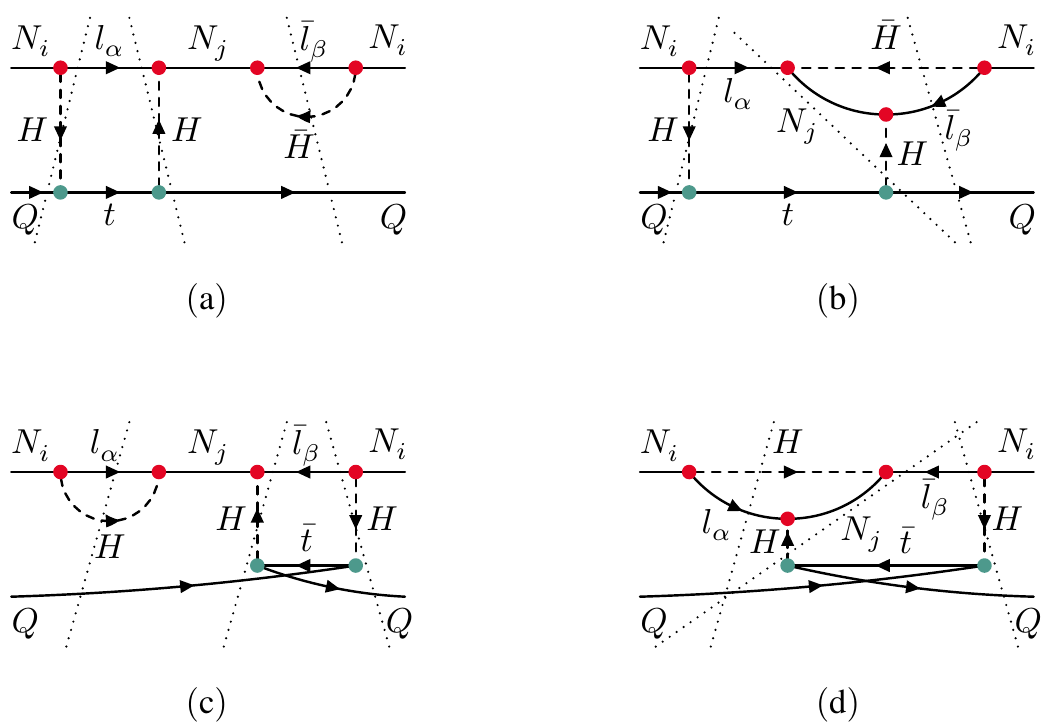}
\caption{Lepton number violating forward-scattering diagrams for the $N_iQ$ initial state contributing to the Higgs thermal-mass effect in $N_i\rightarrow lH$ asymmetry. To obtain a complete list of such diagrams systematically, one can start with vacuum diagrams of the $\mathcal{O}(Y^4Y^2_t)$ order, cutting the $N_i$ and $Q$ propagators in all possible ways \cite{Blazek:2021olf}. In this figure, only the relevant cuts are indicated.}
\label{fig1}
\end{figure}

For the purpose of this work, the relevant contributions of the $\mathcal{O}(Y^4Y^2_t)$ order are generated by the $N_iQ$, $N_i\bar{Q}$, $N_it$, and $N_i\bar{t}$ initial states. We only consider the lepton number violating forward-scattering diagrams  listed in Fig. \ref{fig1} for the $N_iQ$ initial state. These have to be cut several times, respecting the kinematics, following Eq. \eqref{eq1.4}.

Let us examine a specific way of cutting the diagram in Fig. \ref{fig1a}, contributing to the $N_iQ\rightarrow lHQ$ reaction rate asymmetry \cite{Blazek:2021olf}
\begin{align}\label{eq2.2}
\Delta\mathring{\gamma}^{(1a)}_{N_i Q\rightarrow lHQ}=
&\includegraphics[scale=1,valign=c]{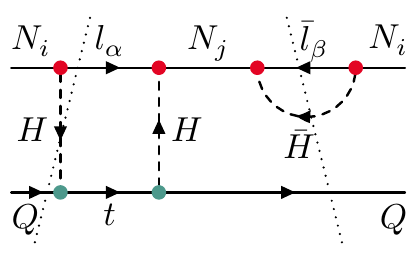} 
+\includegraphics[scale=1,valign=c]{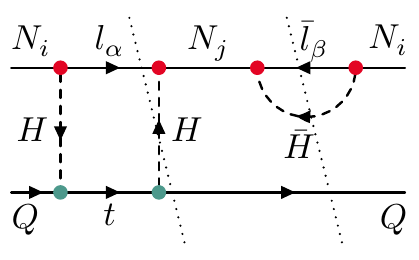}\\
&-\includegraphics[scale=1,valign=c]{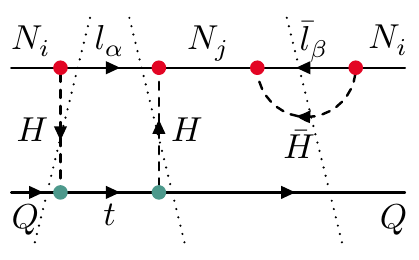} - \hskip2mm\mathrm{m.t.} \nonumber
\end{align}
where the last diagram with three holomorphic cuts comes with a minus sign as indicated in Eq. \eqref{eq1.4}. The $\bar{l}\bar{H}$ loop must be cut to violate $CP$ or, alternatively, to produce the imaginary part of the loop integral. The other diagrams in Figs. \ref{fig1b}-\ref{fig1d} can be cut similarly, contributing to the asymmetry of the same process. In Eq. \eqref{eq2.2}, the $\mathrm{m.t.}$ stands for the abbreviation of the so-called \emph{mirrored terms}. Here they originate in cutting the diagram
\begin{align}\label{eq2.3}
\includegraphics[scale=1,valign=c]{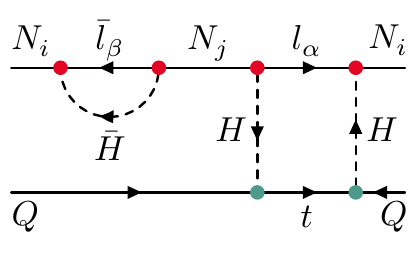}
\end{align}
obtained from Fig. \ref{fig1a} viewed in a mirror with all arrows reversed. No quantum statistics is considered at this place. Thus, no statistical factors are included in the intermediate states, and we only work with circled (i.e. zero-temperature) reaction rates.

Considering the cuts of the box subdiagram appearing in Eq. \eqref{eq2.2}, which can be att\-ri\-bu\-ted to the anomalous thresholds in the $N_iQ$ scattering \cite{Hannesdottir:2022bmo}, two remarkable observations can be made. 

First, putting one of the Higgses on its mass shell immediately puts the second one on the mass shell too, seemingly obtaining an ill-defined expression with delta function squared. Similar issues have been resolved in Refs. \cite{Frye:2018xjj, Racker:2018tzw}, while the present $N_iQ\rightarrow lHQ$ asymmetry has been studied in our previous work in Ref. \cite{Blazek:2021olf}. To overcome this difficulty, we deal with the on-shell intermediate states using the distributional identity
\begin{align}\label{eq2.4}
\frac{1}{k^2+\iu\epsilon} = \mathrm{P.V.}\frac{1}{k^2}-\iu\pi\delta(k^2)\quad\text{or}\quad
\includegraphics[scale=1,valign=c]{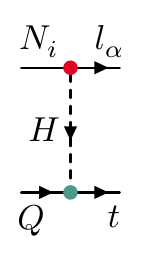} = \mathrm{P.V.}\includegraphics[scale=1,valign=c]{math6a.pdf} + \frac{1}{2}\includegraphics[scale=1,valign=c]{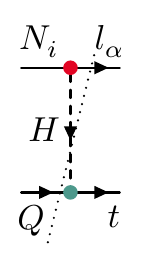}
\end{align}
where only the positive-frequency part of the delta function has been included in the second equality. Applying Eq. \eqref{eq2.4} to the Higgs propagators cancels the last term with three si\-mul\-ta\-ne\-ous cuts in Eq. \eqref{eq2.2}. The two remaining terms are equal, and we obtain \cite{Blazek:2021olf}
\begin{align}\label{eq2.5}
\Delta\mathring{\gamma}^{(1a)}_{N_i Q\rightarrow lHQ}= 2\mathrm{P.V.}\includegraphics[scale=1,valign=c]{math1a_cut2.pdf} - \hskip2mm\mathrm{m.t.}
\end{align}
where the principal value refers to the uncut Higgs line. Now, we can employ \cite{Racker:2018tzw, Frye:2018xjj}
\begin{align}\label{eq2.6}
2\delta_+(k^2)\mathrm{P.V.}\frac{1}{k^2} = 
-\frac{1}{(k^0+\vert\mathbf{k}\vert)^2}\frac{\partial\delta(k^0-\vert\mathbf{k}\vert)}{\partial k^0}
\end{align}
with $\delta_+(k^2) = \theta(k^0)\delta(k^2)$, and, up to a regularized infrared divergence, a well-defined result is obtained.

Second, the cuts in Eq. \eqref{eq2.2} leave the $Q$ momentum unchanged. Therefore, it is reasonable to expect the contribution  to be a correction to the leading-order $N_i\rightarrow lH$ decay asymmetry. Indeed, similar relations, or the so-called \emph{mass-derivative relations}, have been derived for the $CP$ symmetric part of the rate, reading \cite{Blazek:2021gmw}
\begin{align}\label{eq2.7}
\mathring{\gamma}_{N_iQ\rightarrow lHQ\vphantom{\bar{Q}}} + \mathring{\gamma}_{N_i\bar{Q}\rightarrow lH\bar{Q}} + \mathring{\gamma}_{N_it\rightarrow lHt\vphantom{\bar{Q}}} + \mathring{\gamma}_{N_i\bar{t}\rightarrow lH\bar{t}\vphantom{\bar{Q}}}
=\mathring{m}^2_{H,Y_t}(T)\frac{\partial}{\partial m^2_H} \bigg\vert_{m^2_H=0}\mathring{\gamma}_{N_i\rightarrow lH}.
\end{align}
On the left-hand side, the rates  have been obtained by cutting the box diagrams similar to those in Fig. \ref{fig1}, with no additional asymmetry-generating loops. The right-hand side contains the top Yukawa contribution to Higgs thermal mass 
\begin{align}\label{eq2.8}
\mathring{m}^2_{H,Y_t}(T)=12 Y^2_t  \int [d\mathbf{p}_Q] \mathring{f}_Q
\end{align}
and its effect in the $N_i$ decay kinematics. However, the usual definition of the thermal mass includes the uncircled Fermi-Dirac quark density, which is not the case here, and we will come to that later. In Eq. \eqref{eq2.7}, we omitted the thermal part of the wave-function renormalization of the cut Higgs lines, as it vanishes when dimensional regularization is used \cite{Salvio:2011sf}. A more general case can be found in Appendix \ref{appB}. 

One may naturally raise a question -- can a relation similar to that in Eq. \eqref{eq2.7} be formulated for the $N_i\rightarrow lH$ decay asymmetry on the right-hand side? Remarkably, the answer is positive. Let us consider the cuts of the diagrams in Fig. \ref{fig1} as seen in Eq. \eqref{eq2.2}. Summing them up, we obtain the contribution to the circled $N_iQ\rightarrow lHQ$ asymmetry
\begin{align}\label{eq2.9}
\Delta\mathring{\gamma}^{(1a)}_{N_i Q\rightarrow lHQ}+
\Delta\mathring{\gamma}^{(1b)}_{N_i Q\rightarrow lHQ}+
\Delta\mathring{\gamma}^{(1c)}_{N_i Q\rightarrow lHQ}+
\Delta\mathring{\gamma}^{(1d)}_{N_i Q\rightarrow lHQ}=\\
12Y^2_t\int[d\mathbf{p}_{N_i}][d\mathbf{p}_{Q}]\mathring{f}_{N_i}\mathring{f}_{Q}
\sum_{j\neq i} \frac{\Im(Y^\dagger Y)^2_{ij}}{16\pi^2}
\frac{M_j}{M_i}\Bigg[\frac{M^2_i}{M^2_i - M^2_j} - \frac{1}{2}\ln\frac{M^2_i+M^2_j}{M^2_j}\Bigg].\nonumber
\end{align}
To include other initial states with $Q$ replaced by $\bar{Q}$, $t$, and $\bar{t}$, we only need to multiply this result by the factor of four. Unlike the contributions of particular diagrams, listed in Appendix \ref{appA}, the expression in Eq. \eqref{eq2.9} does not depend on the energy squared $s=(p_{N_i}+p_Q)^2$ of the reaction. In fact, the integration over the quark momenta and phase-space density can be factored out, suggesting a possible connection to the Higgs thermal mass due to its interaction with quarks in the thermal medium. 

Next we consider the $N_i\rightarrow lH$ asymmetry evaluated at a finite value of the Higgs mass. We compute its derivative and take the massless limit, obtaining 
\begin{align}\label{eq2.10}
\frac{\partial}{\partial m^2_H}\bigg\vert_{m_H =0} \Delta\mathring{\gamma}_{N_i\rightarrow lH}=&
\int [d\mathbf{p}_{N_i}]\mathring{f}_{N_i}
\sum_{j\neq i} \frac{\Im(Y^\dagger Y)^2_{ij}}{4\pi^2}\\
&\times\frac{M_j}{M_i}
\Bigg[\frac{M^2_i}{M^2_i - M^2_j} - \frac{1}{2}\ln\frac{M^2_i+M^2_j}{M^2_j}\Bigg].\nonumber
\end{align}
The result is four times the result in Eq. \eqref{eq2.9} and we see that the sum of the $N_iQ\rightarrow lHQ$, $N_i{Q}\rightarrow lH\bar{Q}$, $N_it\rightarrow lHt$, and $N_i{t}\rightarrow lH\bar{t}$ asymmetries receives the contribution from the Higgs thermal mass, similarly to Eq. \eqref{eq2.7} for the symmetric part. However, we need to emphasize that the mass-derivative only stands for a part of these asymmetries. There are additional cuttings, namely 
\begin{align}\label{eq2.11}
\Delta\mathring{\gamma}_{N_i(Q)\rightarrow lH(Q)}= \includegraphics[scale=1,valign=c]{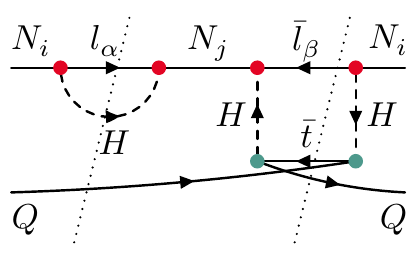}
+\includegraphics[scale=1,valign=c]{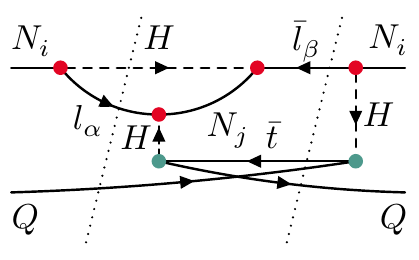}
- \hskip2mm\mathrm{m.t.}
\end{align}
where on the left-hand side we put $Q$ into brackets to distinguish these contributions from the full $N_iQ\rightarrow lHQ$ asymmetry written as
\begin{align}\label{eq2.12}
\Delta\mathring{\gamma}_{N_iQ\rightarrow lHQ\vphantom{)}}=
\Delta\mathring{\gamma}_{N_i(Q)\rightarrow lH(Q)}+
\frac{1}{4}\mathring{m}^2_{H,Y_t}(T)\frac{\partial}{\partial m^2_H} \bigg\vert_{m^2_H=0}\Delta\mathring{\gamma}_{N_i\rightarrow lH}.
\end{align}

\section{Higher-order unitarity constraints at finite temperature}\label{sec3}

We may have noticed that each diagram in Eq. \eqref{eq2.11} comes with two left-handed quarks in an identical state -- a contribution that may seem nonsense. Nevertheless, we must include those terms to achieve infrared finiteness \cite{Frye:2018xjj, Blazek:2021olf}. To better understand their physical origin, let us compare the cuts of a higher-order diagram contributing to the $N_i\rightarrow lH$ asymmetry
\begin{align}\label{eq3.1}
\includegraphics[scale=1,valign=c]{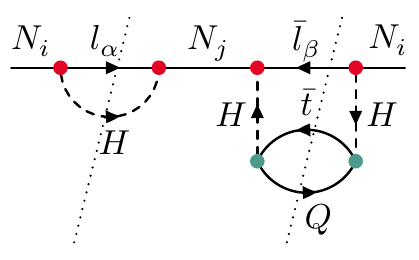} - \hskip2mm\mathrm{m.t.}
\end{align}
to the first term in Eq. \eqref{eq2.11}. Up to a relative minus sign, the only difference between the two contributions is the presence of the $\mathring{f}_Q$ in the $N_iQ\rightarrow lHQ$ asymmetry due to the definition of the circled rates in Eq. \eqref{eq1.5}. In the previous section, the Higgs thermal mass entered via anomalous thresholds shown in Fig. \ref{fig1}. Here, we observe that including the diagrams in Eq. \eqref{eq2.11} brings quantum statistics in. Considering higher winding numbers of the $Q$ line, cutting them accordingly, and summing the geometric series of Eq. \eqref{eq1.12} eventually generates a complete $1-f_Q$ statistical factor in the respective cut. Next, we can open the same $Q$ line in Eq. \eqref{eq3.1} without crossing, obtaining the $N_i\bar{Q}\rightarrow lH\bar{Q}$ asymmetry contribution
\begin{align}\label{eq3.2}
\includegraphics[scale=1,valign=c]{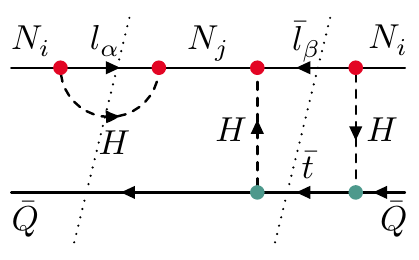} - \hskip2mm\mathrm{m.t.}
\end{align}
Let us add the cuts of the diagrams in Eqs. \eqref{eq2.11} and \eqref{eq3.2} together, including the $Q$ line windings. The result can be seen as the zero-temperature $Q$ propagator in Eq. \eqref{eq3.1} extended with an extra term that corresponds to the medium contribution \cite{Das:1997gg}
\begin{align}\label{eq3.3}
\frac{\iu}{p_Q^2+\iu\epsilon}\quad\rightarrow\quad
\frac{\iu}{p_Q^2+\iu\epsilon} - 2\pi f^{\vphantom{0}}_{Q}(\vert p^0_Q\vert)\delta(p^2_Q).
\end{align}
This observation uncovers the true nature of the cuttings in Eq. \eqref{eq2.11} necessary for infrared finiteness \cite{Blazek:2021olf, Frye:2018xjj}. \emph{The appearance of quantum statistics implies that the naive separation into a quantum description of the interactions and classical treatment of a many-body system does not survive the inclusion of higher-order perturbative corrections.} In fact, when considering the cuttings in Eq. \eqref{eq2.11} using the classical-rate recipe of Eq. \eqref{eq1.5}, we were computing quantum thermal corrections without knowing it. This turns out to be a consequence of the density matrix evolution, sketched in Eq. \eqref{eq1.11} and discussed in the last section of Ref. \cite{Blazek:2021zoj} in more detail.

\subsection{Mass-derivative and quantum statistics}

\begin{figure}
\subfloat{\label{fig2a}}
\subfloat{\label{fig2b}}
\subfloat{\label{fig2c}}
\subfloat{\label{fig2d}}
\centering
\includegraphics[scale=1]{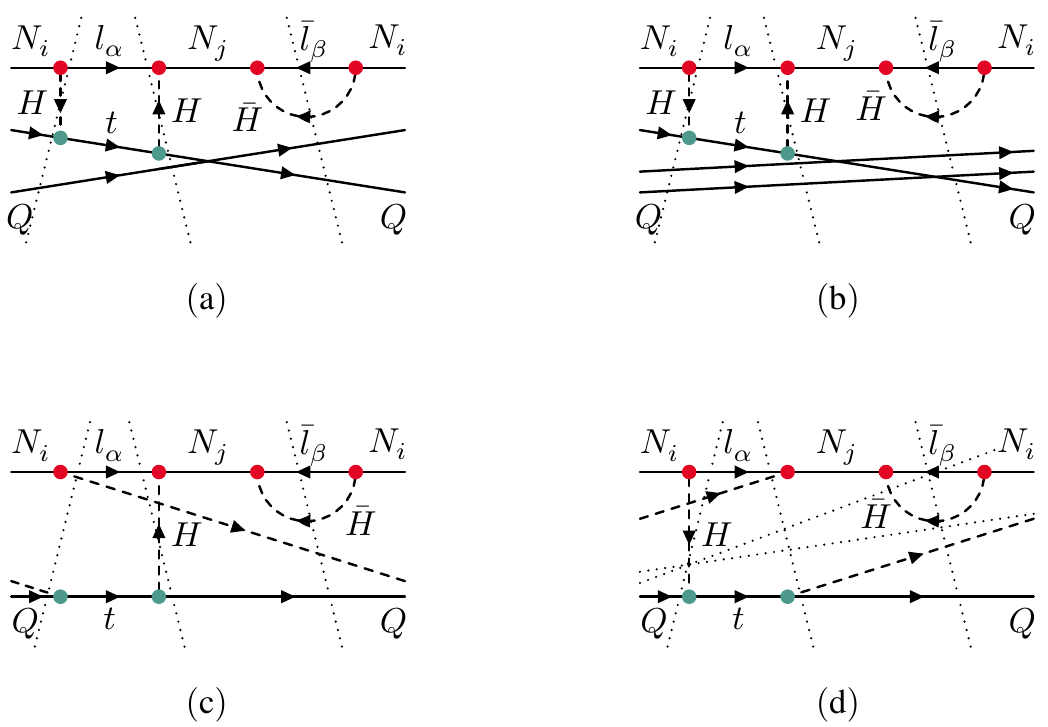}
\caption{Diagrams obtained from Fig. \ref{fig1a} considering $w=1$ and $w=2$ for the $Q$ initial-state line (Figs. \ref{fig2a} and \ref{fig2b}), and $w=1$ for the Higgs propagators (Figs. \ref{fig2c} and \ref{fig2d}). The relevant cuts are indicated, however, not all of them can be made simultaneously in Fig. \ref{fig2d}.}
\label{fig2}
\end{figure}

Turning our attention back to the mass-derivative from the classical rate asymmetries in Eq. \eqref{eq2.11}, we can generalize the procedure to include quantum densities and statistical factors. The uncircled $N_iQ\rightarrow lHQ$ asymmetry receives the contribution from Eq. \eqref{eq2.2} and the diagrams obtained by winding the propagators and external lines. Usually, the winding does not lead to a surprise -- only powers of circled densities are produced. This can be observed in Figs. \ref{fig2a} and \ref{fig2b}, where $w=1$ and $w=2$ cases for the external $Q$ line are shown, respectively. In both these cut diagrams, the $S$-matrix elements are the same as in Fig. \ref{fig1a}.

However, we need to check if steps from Eq. \eqref{eq2.2} to Eq. \eqref{eq2.5} can be reproduced for nonzero winding numbers of the Higgs lines, as shown in Figs. \ref{fig2c} and \ref{fig2d}. These two diagrams formally contribute to the circled $N_i HQ\rightarrow lHHQ$ asymmetry. Summing over all winding numbers, the uncircled $N_iQ\rightarrow lHQ$ asymmetry contribution will be obtained. Cutting the diagram in Fig. \ref{fig2c} according to Eq. \eqref{eq1.4} leads to
\begin{align}\label{eq3.4}
\Delta\mathring{\gamma}^{(2c)}_{N_i HQ\rightarrow lHHQ}=&
\includegraphics[scale=1,valign=c]{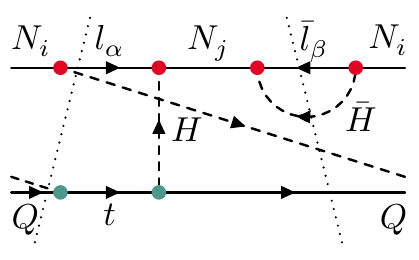}+
\includegraphics[scale=1,valign=c]{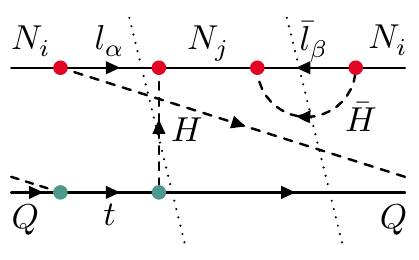}\\
&-\includegraphics[scale=1,valign=c]{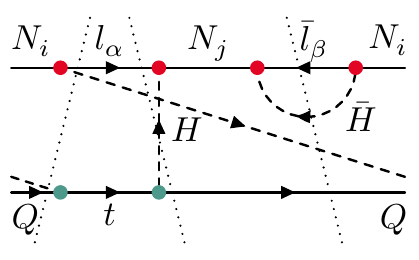} - \hskip2mm\mathrm{m.t.}\nonumber
\end{align}
where, unlike in Eq. \eqref{eq2.2}, the second and third diagrams cancel each other. We are thus left with the first term only, in which the second Higgs line remains uncut. 

The diagram in Fig. \ref{fig2d} can be cut in more distinct ways, that, after several cancelations, can be reduced to
\begin{align}\label{eq3.5}
\Delta\mathring{\gamma}^{(2d)}_{N_i HQ\rightarrow lHHQ}=
\includegraphics[scale=1,valign=c]{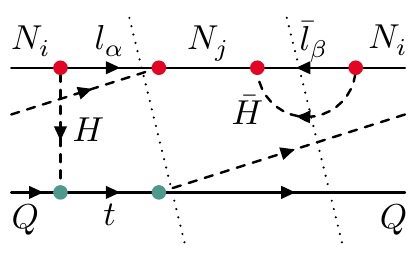}
-\includegraphics[scale=1,valign=c]{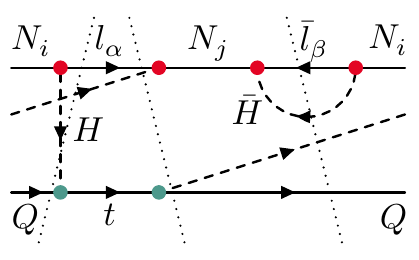}- \hskip2mm\mathrm{m.t.}
\end{align}
We may finally add Eq. \eqref{eq3.5} to the remaining first diagram of Eq. \eqref{eq3.4}. Employing Eq. \eqref{eq2.4} for the uncut Higgs propagators, the diagram with three cuts is canceled, resulting in the expression
\begin{align}\label{eq3.6}
\Delta\mathring{\gamma}^{(2c)}_{N_i HQ\rightarrow lHHQ}+ \Delta\mathring{\gamma}^{(2d)}_{N_i HQ\rightarrow lHHQ}=
2\mathrm{P.V.}\includegraphics[scale=1,valign=c]{math1a_wH1_2cut11.pdf}- \hskip2mm\mathrm{m.t.}
\end{align}
with the same structure as seen in Eq. \eqref{eq2.5}. Again, compared to the lowest-winding-number diagram, only an extra factor of $\mathring{f}_H$ appears in Eq. \eqref{eq3.6}. 

The reader may verify, perhaps after some practice in drawing and cutting, that the same happens for any set of winding numbers. The inclusion of all windings then effectively leads to the replacement of the initial-state circled densities
\begin{align}\label{eq3.7}
\mathring{f}_{N_i\vphantom{\bar{H}}} \mathring{f}_{Q\vphantom{\bar{H}}} 
\quad\rightarrow\quad
f_{N_i\vphantom{\bar{H}}}f_{Q\vphantom{\bar{H}}} (1+f_{H\vphantom{\bar{H}}})(1-f_{l\vphantom{\bar{H}}}) (1+f_{\bar{H}})(1-f_{\bar{l}})
\end{align}
made in Eq. \eqref{eq2.5}. The uncircled asymmetry derived from Fig. \ref{fig1a} can now be written as
\begin{align}\label{eq3.8}
\Delta\gamma^{(1a)}_{N_iQ\rightarrow lHQ}=&12Y^2_t\int[d\mathbf{p}_{N_i}] [d\mathbf{p}_Q]f_{N_i}f_Q \sum_{j\neq i} \frac{\Im(Y^\dagger Y)^2_{ij}}{(2\pi)^{d-2}}
\frac{2M_i M_j}{M^2_i-M^2_j}\\ 
&\times\mu^{4-d}\int d^d k \,\mathcal{F}(k^0,\mathbf{k})\,
2\delta_+(k^2)\mathrm{P.V.}\frac{1}{k^2}\nonumber
\end{align}
where $k$ denotes the Higgs four-momentum, $\mu$ is a mass scale introduced to preserve the correct dimension of the expression, and
\begin{align}\label{eq3.9}
\mathcal{F}(k^0,\mathbf{k})=& \delta_+[(p_{N_i}-k)^2]
[1+f_H(k^0)][1-f_l(E_{N_i}-k^0)]\frac{2p_Q.k}{2p_Q.k+k^2}\\
&\times\mu^{4-d}
\int[d\mathbf{p}_{\bar{H}}][d\mathbf{p}_{\bar{l}}]\, (2\pi)^d
\delta^{(d)}(p_{N_i\vphantom{\bar{H}}}-p_{\bar{l}}-p_{\bar{H}})\,
(p_{\bar{H}}-k)^2(1+f_{\bar{H}})(1-f_{\bar{l}}).\nonumber
\end{align}
Here the energy dependence of the phase-space densities is only explicit when necessary. Further inclusion of uncircled asymmetries from cuttings of the diagrams in Figs. \ref{fig1b}-\ref{fig1d} and their windings extends function $\mathcal{F}(k^0,\mathbf{k})$ to a more complicated expression proportional to
\begin{align}\label{eq3.10}
\frac{2p_Q.k}{2p_Q.k+k^2}+\frac{2p_Q.k}{2p_Q.k-k^2}.
\end{align}
Employing the identity in Eq. \eqref{eq2.6} and \cite{Blazek:2021gmw}
\begin{align}\label{eq3.11}
\frac{\partial}{\partial k^0}\bigg\vert_{k^0=\vert\mathbf{k}\vert} \frac{\mathcal{F}(k^0,\mathbf{k})}{(k^0+\vert\mathbf{k}\vert)^2}=
\frac{\partial}{\partial m^2_H} \bigg\vert_{m_H=0}\frac{\mathcal{F}(E_{\mathbf{k}},\mathbf{k})}{2E_{\mathbf{k}}}
\quad\text{for}\quad E_{\mathbf{k}}=\sqrt{m^2_H+\mathbf{k}^2}
\end{align}
the result can be transformed into a mass-derivative. The expression in Eq. \eqref{eq3.10} vanishes when derived as in Eq. \eqref{eq3.11}, and may be replaced by an overall factor of two. The rest will not depend on the $Q$ momentum, and thermal mass \cite{Comelli:1996vm, Giudice:2003jh}
\begin{align}\label{eq3.12}
m^2_{H,Y_t}(T)=12 Y^2_t  \int [d\mathbf{p}_Q] f_Q = \frac{1}{4}Y^2_t T^2
\end{align}
can be factored out. Finally, we can write
\begin{align}\label{eq3.13}
\Delta\gamma_{N_iQ\rightarrow lHQ\vphantom{)}}=
\Delta\gamma_{N_i(Q)\rightarrow lH(Q)}+
\frac{1}{4}m^2_{H,Y_t}(T)\frac{\partial}{\partial m^2_H} \bigg\vert_{m^2_H=0}\Delta\gamma_{N_i\rightarrow lH}
\end{align}
with quantum-corrected asymmetries, in which the statistical factors in the intermediate states are included.

\subsection{$N_iQ$ asymmetries and unitarity constraints}

From the general considerations of the first section, we may expect the sum of all equilibrium asymmetries with the $N_iQ$ initial state to vanish. However, Eq. \eqref{eq1.4} offers a more powerful tool, as we can now track the asymmetry cancelations in pairs. 

The equilibrium $N_iQ\rightarrow lHQ$ asymmetry from the diagram in Fig. \ref{fig1a}, shown in Eq. \eqref{eq2.2}, is canceled by the part of the $N_iQ\rightarrow \bar{l}\bar{H}Q$ asymmetry originating from the mirrored diagram in Eq. \eqref{eq2.3}. The roles of the original and mirrored terms in these two contributions are exchanged, and thus, they come with opposite signs. Similarly, both diagrams in Figs. \ref{fig1a} and \ref{fig1b} lead to the $N_iQ\rightarrow lt$ rate asymmetry from
\begin{align}\label{eq3.14}
\mathring{\gamma}_{N_iQ\rightarrow lt} =
\includegraphics[scale=1,valign=c]{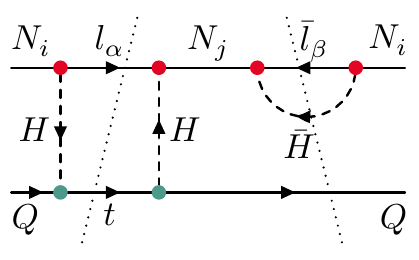}
+\includegraphics[scale=1,valign=c]{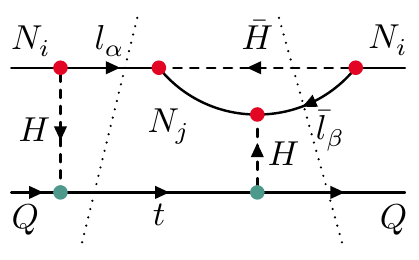}- \hskip2mm\mathrm{m.t.}
\end{align}
canceled by the $N_i(Q)\rightarrow \bar{l}\bar{H}(Q)$ contribution from the respective mirrored diagrams. Combining everything together, we can derive the $CPT$ and unitarity constraints \cite{Blazek:2021olf}
\begin{align}\label{eq3.15}
\Delta\mathring{\gamma}_{N_iQ\rightarrow lt\vphantom{\bar{l}}}+
\Delta\mathring{\gamma}_{N_iQ\rightarrow lHQ\vphantom{\bar{l}}}+
\Delta\mathring{\gamma}_{N_iQ\rightarrow \bar{l}\bar{H}Q}+
\Delta\mathring{\gamma}_{N_iQ\rightarrow \bar{l}QQ\bar{t}} = 0.
\end{align}
Here the asymmetries, even though obtained within the classical kinetic theory, come with thermal-mass effects and a quantum statistical factor from the $N_i\rightarrow \bar{l}Q\bar{t}$ decay asymmetry. This factor is obtained from
\begin{align}
\mathring{\gamma}_{N_iQ\rightarrow \bar{l}QQ\bar{t}} =
\includegraphics[scale=1,valign=c]{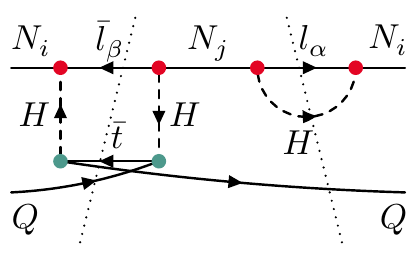}
+\includegraphics[scale=1,valign=c]{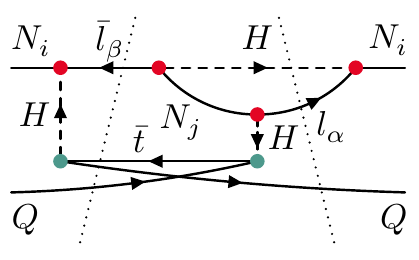} - \hskip2mm\mathrm{m.t.}
\end{align}
with the mirrored terms shown in Eq. \eqref{eq2.11}. The remarkable appearance of quantumness in higher-order $CPT$ and unitarity constraints, earmarked in the abstract, is explicit now.

The uncircled asymmetries differ from the circled ones by the replacements similar to Eq. \eqref{eq3.7}. They can be transformed into infinite sequences of zero-temperature contributions of particular winding numbers in each forward-scattering diagram. For each set of winding numbers, a relation similar to Eq. \eqref{eq3.14} can be derived. Therefore, the uncircled asymmetries obey
\begin{align}\label{eq3.17}
\Delta\gamma_{N_iQ\rightarrow lt\vphantom{\bar{l}}}+
\Delta\gamma_{N_i(Q)\rightarrow lH(Q)\vphantom{\bar{l}}}+
\Delta\gamma_{N_i(Q)\rightarrow \bar{l}\bar{H}(Q)}+
\Delta\gamma_{N_iQ\rightarrow \bar{l}QQ\bar{t}} = 0
\end{align}
where we used Eq. \eqref{eq3.13} to separate the mass-derivative of the leading-order part
\begin{align}\label{eq3.18}
\Delta\gamma_{N_i\rightarrow lH\vphantom{\bar{l}}}+
\Delta\gamma_{N_i\rightarrow \bar{l}\bar{H}}
\end{align}
vanishing independently. The second and third terms in Eq. \eqref{eq3.17} correspond to the $f_Q$ part of the $1-f_Q$ statistical factor appearing in the top-Yukawa corrections to the $CP$ violating loops in the $N_i\rightarrow lH$ and $N_i\rightarrow \bar{l}\bar{H}$ decay asymmetries, similar to Eq. \eqref{eq2.11} for the lowest-winding number case. In equilibrium, these are canceled by the $N_iQ\rightarrow lt$ asymmetry together with a part the final-state statistical factor in the $N_i\rightarrow \bar{l}Q\bar{t}$ asymmetry.

\section{Summary}

In this work, we have derived the $CPT$ and unitarity constraints for higher-order $CP$ asy\-mme\-tries within the seesaw type-I leptogenesis model. The right-handed neutrino and left-handed top quark initial states have been used as an example. The inclusion of quantum thermal corrections makes tracking equilibrium asymmetry cancelations difficult, as the on-shell parts of the propagators, which are essential to the asymmetry generation, are modified by quantum statistical factors. The obstacles have been overcome in two stages. First, we employed the holomorphic cutting rules \cite{Coster:1970jy, Bourjaily:2020wvq, Hannesdottir:2022bmo}, which allow for a diagrammatic representation of $CP$ asymmetries at higher perturbative orders \cite{Blazek:2021olf}. Then, cylindrical diagrams with all possible winding numbers of internal lines have been cut to systematically account for quantum thermal effects in reaction rates and their asymmetries \cite{Blazek:2021zoj}. Finally, the $CPT$ and unitarity constraints, including thermal-mass effects and quantum statistics in both final states and asymmetry-generating loops, have been presented.

\acknowledgments
Tom\'{a}\v{s} Bla\v{z}ek and Peter Mat\'{a}k were supported by the Slovak Ministry of Education Contract No. 0466/2022. Viktor Zaujec received funding from the Comenius University in Bratislava, grant No. UK/317/2022: \emph{Unitarita, CP asymetrie a diagramatick\'{a} reprezent\'{a}cia tepeln\'{y}ch efektov v leptogen\'{e}ze}. We also would like to thank our colleague Fedor \v{S}imkovic for his long-term support.

\appendix

\section{Circled asymmetries from the $N_iQ$ anomalous thresholds}\label{appA}

In this appendix, we present the explicit results for the asymmetry in Eq. \eqref{eq2.2} and analogous cuts of the diagrams in Figs. \ref{fig1b}-\ref{fig1d}. To regularize infrared divergences, we can perform the calculation in $d$ dimensions, leading to
\begin{align}\label{A.1}
\Delta\mathring{\gamma}^{(1a)}_{N_i Q\rightarrow lHQ}= &12Y^2_t\int[d\mathbf{p}_{N_i}][d\mathbf{p}_{Q}]\mathring{f}_{N_i}\mathring{f}_{Q}
\frac{\Lambda^2_d}{(4\pi)^{d-2}} \sum_{j\neq i}\Im(Y^\dagger Y)^2_{ij}\\
&\times\frac{M_iM_j}{M^2_i-M^2_j}\bigg[d-2+\frac{1}{d-4}\frac{2M^2_i}{s-M^2_i}\bigg]
\nonumber
\end{align}
where $s=(p_{N_i}+p_Q)^2$ and
\begin{align}\label{A.2}
\Lambda_d = \frac{1}{2}\bigg(\frac{M_i}{\mu}\bigg)^{d-4}\frac{\Gamma(d/2-1)}{\Gamma(d-3)}
\end{align}
with a scale $\mu$ fixing the dimension. For the rest of the $N_iQ\rightarrow lHQ$ asymmetry contributions, the second line in Eq. \eqref{A.1} has to be replaced by
\begin{align}
\frac{M_j}{M_i}\bigg[-1+\frac{M^2_j}{M^2_i}\ln\bigg(1+\frac{M^2_i}{M^2_j}\bigg)+
\bigg(d-3+\frac{1}{d-4}\frac{2M^2_i}{s-M^2_i}\bigg) F(M^2_j/M^2_i)\bigg]\label{A.3}\\
\frac{M_iM_j}{M^2_i-M^2_j}\bigg[d-2-\frac{1}{d-4}\frac{2M^2_i}{s-M^2_i} \bigg]\label{A.4}\\
\frac{M_j}{M_i}\bigg[-1+\frac{M^2_j}{M^2_i}\ln\bigg(1+\frac{M^2_i}{M^2_j}\bigg)+
\bigg(d-3-\frac{1}{d-4}\frac{2M^2_i}{s-M^2_i}\bigg) F(M^2_j/M^2_i)\bigg]\label{A.5}
\end{align}
to reproduce the asymmetries from Figs. \ref{fig1b}, \ref{fig1c}, and \ref{fig1d}, respectively.
The auxiliary function $F(M^2_j/M^2_i)$ is defined as
\begin{align}\label{A.6}
&F(x)=1-(1+x)\ln(1+1/x) 
+\mathcal{O}(d-4).
\end{align}
Using Eqs. \eqref{A.1}-\eqref{A.5}, we may verify that in four dimensions, the resulting asymmetry in Eq. \eqref{eq2.9} is obtained.

\section{Infrared singularities and finite quark mass}\label{appB}

Another way to deal with infrared singularities is to assign a small mass $m$ to $Q$ and $t$ particles as in Refs. \cite{Racker:2018tzw, Blazek:2021gmw}. In that case, the thermal part of the Higgs wave-function renormalization does not vanish, and we need to modify the mass-derivative relations in Eqs. \eqref{eq2.12} and \eqref{eq3.13}. In analogy to the zero-temperature case, we define a circled Higgs self-energy
\begin{align}\label{B.1}
-\iu\mathring{\Pi}_T(k^0,\mathbf{k}) = 
\includegraphics[scale=1,valign=c]{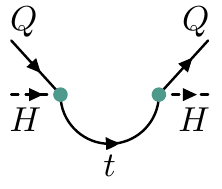}+\includegraphics[scale=1,valign=c]{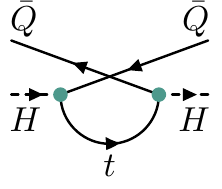}+
\includegraphics[scale=1,valign=c]{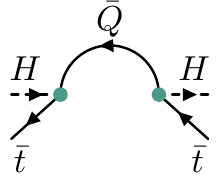}+\includegraphics[scale=1,valign=c]{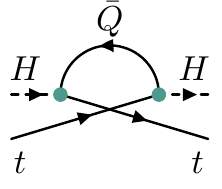}.
\end{align}
where each quark or antiquark line includes momentum integration with circled density and summation over the discrete degrees of freedom. Correspondingly, the thermal mass and wave-function renormalization factor \cite{Blazek:2021gmw}
\begin{align}
\mathring{m}^2_{H,Y_t}(T)=&\mathring{\Pi}_T(k^0,\mathbf{k})\vert_{k^0=\vert\mathbf{k}\vert}=
12Y^2_t\int[d\mathbf{p}_Q]\mathring{f}_Q\label{B.2}\\
\delta\mathring{Z}_{H,Y_t}(T)=&\frac{1}{2\vert\mathbf{k}\vert}\frac{\partial}{\partial k^0}\bigg\vert_{k^0=\vert\mathbf{k}\vert}\Re\mathring{\Pi}_T(k^0,\mathbf{k})=
-6Y^2_t\int[d\mathbf{p}_Q]\mathring{f}_Q\frac{m^2}{(k.p_Q)^2}\bigg\vert_{k^0=\vert\mathbf{k} \vert}\label{B.3}
\end{align}
can be introduced, while the full thermal quantities are obtained via the inclusion of all higher winding numbers of the quark lines. The leading-order $N_i\rightarrow lH$ decay asymmetry
\begin{align}\label{B.4}
\Delta\mathring{\gamma}_{N_i\rightarrow lH}=
\includegraphics[scale=1,valign=c]{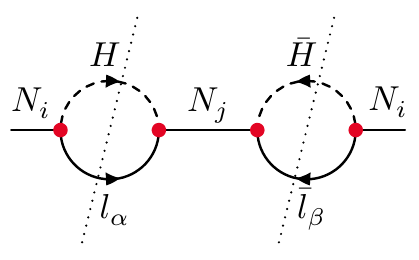}+\includegraphics[scale=1,valign=c]{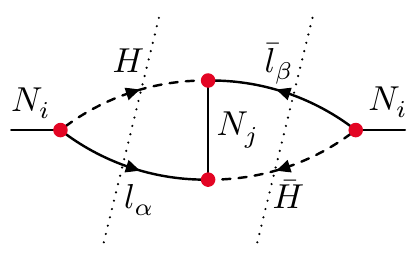}
- \hskip2mm\mathrm{m.t.}
\end{align}
results from the diagrams with two scalar (Higgs or anti-Higgs) cut lines. In each of them and from each side of the cut, the factor of $\delta\mathring{Z}_{H,Y_t}(T)/2$ adds to the $\mathcal{O}(Y^4Y^2_t)$ asymmetries
\begin{align}\label{B.5}
\Delta\mathring{\gamma}^{\delta Z}_{N_i\rightarrow lH} = &12Y^2_t\int[d\mathbf{p}_{N_i}][d\mathbf{p}_{Q}]\mathring{f}_{N_i}\mathring{f}_{Q}
\sum_{j\neq i} \frac{\Im(Y^\dagger Y)^2_{ij}}{4\pi^2}\\
&\times\frac{M_j}{M_i}\Bigg[\frac{M^2_i}{M^2_i-M^2_j}+1-
\bigg(1+\frac{M^2_j}{M^2_i}\bigg)\ln\bigg(1+\frac{M^2_i}{M^2_j}\bigg) \Bigg].\nonumber
\end{align}
The contributions from Figs. \ref{fig1a}-\ref{fig1d} are now equal to
\begin{align}
\Delta\mathring{\gamma}^{(1a)}_{N_i Q\rightarrow lHQ}=& 3Y^2_t\int[d\mathbf{p}_{N_i}][d\mathbf{p}_{Q}]\mathring{f}_{N_i}\mathring{f}_{Q}\sum_{j\neq i}\frac{\Im(Y^\dagger Y)^2_{ij}}{16\pi^2} \label{B.6}\\
&\times\frac{M_iM_j}{M^2_i-M^2_j}
\bigg[1+\frac{s}{M^2_i}-\frac{M^2_i}{s-M^2_i}
\bigg(\ln\frac{M^2_i}{s-M^2_i}+\ln\frac{m}{M_i}\bigg)\bigg]
\nonumber\\
\Delta\mathring{\gamma}^{(1b)}_{N_i Q\rightarrow lHQ}=& 3Y^2_t\int[d\mathbf{p}_{N_i}][d\mathbf{p}_{Q}]\mathring{f}_{N_i}\mathring{f}_{Q}\sum_{j\neq i}\frac{\Im(Y^\dagger Y)^2_{ij}}{16\pi^2}
\frac{M_j}{M_i}
\bigg[-\frac{1}{2}\ln\bigg(1+\frac{M^2_i}{M^2_j}\bigg)\label{B.7}\\
&+\bigg(\frac{s}{M^2_i}-\frac{M^2_i}{s-M^2_i}\bigg(\ln\frac{M^2_i}{s-M^2_i}+ \ln\frac{m}{M_i}\bigg)\bigg)F(M^2_j/M^2_i)\bigg]
\nonumber\\
\Delta\mathring{\gamma}^{(1c)}_{N_i Q\rightarrow lHQ}=& 3Y^2_t\int[d\mathbf{p}_{N_i}][d\mathbf{p}_{Q}]\mathring{f}_{N_i}\mathring{f}_{Q}\sum_{j\neq i}\frac{\Im(Y^\dagger Y)^2_{ij}}{16\pi^2} \label{B.8}\\
&\times\frac{M_iM_j}{M^2_i-M^2_j}
\bigg[3-\frac{s}{M^2_i}+\frac{M^2_i}{s-M^2_i}\bigg(
\ln\frac{M^2_i}{s-M^2_i}+\ln\frac{m}{M_i}\bigg)\bigg]\nonumber\\
\Delta\mathring{\gamma}^{(1d)}_{N_i Q\rightarrow lHQ}=& 3Y^2_t\int[d\mathbf{p}_{N_i}][d\mathbf{p}_{Q}]\mathring{f}_{N_i}\mathring{f}_{Q}\sum_{j\neq i}\frac{\Im(Y^\dagger Y)^2_{ij}}{16\pi^2}
\frac{M_j}{M_i}\bigg[-\frac{1}{2}\ln\bigg(1+\frac{M^2_i}{M^2_j}\bigg) \label{B.9}\\
&+\bigg(2-\frac{s}{M^2_i}+\frac{M^2_i}{s-M^2_i}\bigg(\ln\frac{M^2_i}{s-M^2_i}+ \ln\frac{m}{M_i}\bigg)\bigg) F(M^2_j/M^2_i)\bigg]\nonumber
\end{align}
and their sum, independent of $m$, is infrared finite. We can easily observe that it includes both the mass-derivative and the wave-function renormalization, as shown in Eq. \eqref{B.5}. The formula in Eq. \eqref{eq2.12} thus can be generalized to
\begin{align}
\Delta\mathring{\gamma}_{N_iQ\rightarrow lHQ\vphantom{)}}=
\Delta\mathring{\gamma}_{N_i(Q)\rightarrow lH(Q)}+
\frac{1}{4}\mathring{m}^2_{H,Y_t}(T)\frac{\partial}{\partial m^2_H} \bigg\vert_{m^2_H=0}\Delta\mathring{\gamma}_{N_i\rightarrow lH}+ \frac{1}{4}\Delta\mathring{\gamma}^{\delta Z}_{N_i\rightarrow lH}
\end{align}
and similarly for the uncircled rates.

\bibliographystyle{JHEP.bst}
\bibliography{CLANOK.bib}

\end{document}